\documentclass[paper]{ieice}
\usepackage[dvips]{graphicx}
\usepackage[fleqn]{amsmath}
\usepackage {amssymb,amsthm,amsbsy}
\usepackage[varg]{txfonts}
\usepackage{enumerate}
\field{A}
\title{On the Hamming Auto- and Cross-correlation  Functions of a Class of Frequency Hopping Sequences of Length $ p^{n} $ }
\authorlist{
\authorentry[mlqiecully@163.com]{Minglong Qi}{m}{etab1}
\authorentry{Shengwu Xiong}{n}{etab1}
\authorentry{Jingling Yuan}{n}{etab1}
\affiliate[etab1]{School of Computer Science and Technology, Wuhan University of Technology, Mafangshan West Campus, 430070 Wuhan City, China }
}
\theoremstyle{definition}
\newtheorem{sec1_thm1}{Theorem}[section]

\newtheorem{sec1_lemma1}{Lemma}[section]
\newtheorem{sec1_lemma2}[sec1_lemma1]{Lemma}
\newtheorem{sec1_lemma3}[sec1_lemma1]{Lemma}

\newtheorem{sec1_cor1}{Corollary}[section]

\newtheorem{sec1_def1}{Definition}[section]
\newtheorem{sec1_def2}[sec1_def1]{Definition}
\newtheorem{sec1_def3}[sec1_def1]{Definition}
\newtheorem{sec1_def4}[sec1_def1]{Definition}

\newtheorem{sec2_lemma1}{Lemma}[section]
\newtheorem{sec2_lemma2}[sec2_lemma1]{Lemma}

\newtheorem{sec2_thm1}{Theorem}[section]
\newtheorem{sec2_thm2}[sec2_thm1]{Theorem}

\newtheorem{sec2_cons1}{Construct.}[section]

\newtheorem{sec3_prop1}{Proposition}[section]
\newtheorem{sec3_prop2}[sec3_prop1]{Proposition}
\newtheorem{sec3_prop3}[sec3_prop1]{Proposition}

\newtheorem{sec3_thm1}{Theorem}[section]

\newtheorem{sec3_cor1}{Corollary}[section]

\newtheorem{sec4_thm1}{Theorem}[section]

\begin{document}
\maketitle
\begin{summary}
In this paper, a new class of frequency hopping sequences (FHSs) of length $  p^{n} $ is constructed by using Ding-Helleseth generalized cyclotomic classes of order 2, of which the Hamming auto- and cross-correlation  functions are investigated (for the Hamming cross-correlation, only the case $ p\equiv 3\pmod 4 $ is considered). It is shown that the set of the  constructed FHSs is optimal with respect to the average Hamming correlation functions.
\end{summary}

\section{Introduction}

Let $ \mathcal{F}=\lbrace f_{0},f_{1},\cdots,f_{m-1}\rbrace $ be a set of $ m $ elements called the alphabet of available frequencies.  A sequence with $ \nu $ elements taken from $ \mathcal{F} $ is said to be a frequency hopping sequence (FHS) over $ \mathcal{F} $ of length $ \nu $. Let $ \mathbf{X} $, $ \mathbf{Y} $ be two  FHSs taken from a set with $ M $ FHSs, $ \mathcal{S} $, i. e. , $ \mathbf{X}=\left(\mathbf{X}(t)\right)_{t=0}^{\nu -1} $,  $ \mathbf{Y}=\left(\mathbf{Y}(t)\right)_{t=0}^{\nu -1} $ 
where $ \mathbf{X}(t) $, $ \mathbf{Y}(t) \in \mathcal{F} $,  $ 0\le t\le \nu -1 $.
Define the periodic Hamming cross-correlation function between $ \mathbf{X} $ and $ \mathbf{Y} $ as the following equation:
\begin{equation}\label{hamming_cross_cor}
H(\mathbf{X},\mathbf{Y}:\tau)=\sum_{t=0}^{\nu-1}h[\mathbf{X}(t+\tau),\mathbf{Y}(t)], 0\le \tau <\nu
\end{equation}
where $  h[\mathbf{X}(t+\tau),\mathbf{Y}(t)]=1 $ if  $  \mathbf{X}(t+\tau)=\mathbf{Y}(t) $ , and $ 0 $ otherwise. The subscript $ t+\tau $ in (\ref{hamming_cross_cor}) is performed  modulo  $ \nu $.

 Set $ \mathbf{Y}=\mathbf{X} $ in  (\ref{hamming_cross_cor}), then $ H(\mathbf{X},\mathbf{X}:\tau) $ with $ 0<\tau<\nu $ is called  the Hamming autocorrelation function  of $  \mathbf{X} $, denoted by $ H(\mathbf{X}:\tau) $.
 
 If the FHSs set, $ \mathcal{S} $, is explicitly enumarated as  $ \mathcal{S}=\lbrace \mathbf{X}_{0},\mathbf{X}_{1},\cdots,\mathbf{X}_{M-1}\rbrace $, then we use $ H(i,j:\tau) $ to denote the Hamming cross-correlation function between $ \mathbf{X}_{i} $ and $ \mathbf{X}_{j} $, and $ H(i:\tau) $ to denote the Hamming autocorrelation function of $ \mathbf{X}_{i} $, where $0\le i,j<M $.
 
 We need some maximum parameters on the FHSs in order to describe  two important theoretical bounds described in the sequel. Let $ \mathbf{X},\mathbf{Y} \in \mathcal{S} $. Define
 \begin{equation*}\label{haming_max}
 \begin{split}
 H(\mathbf{X})&=\max\limits_{1\le\tau<\nu}\lbrace H(\mathbf{X}:\tau)\rbrace,\\
  H(\mathbf{X},\mathbf{Y})&=\max\limits_{0\le\tau<\nu,\mathbf{X}\ne \mathbf{Y}}\lbrace H(\mathbf{X},\mathbf{Y}:\tau)\rbrace,\\
  H(\mathcal{S})&=\max\lbrace \max\limits_{\mathbf{X}\in\mathcal{S}}\lbrace H(\mathbf{X}) \rbrace,\max\limits_{\mathbf{X},\mathbf{Y}\in\mathcal{S},\mathbf{X}\ne \mathbf{Y}}\lbrace H(\mathbf{X},\mathbf{Y})\rbrace\rbrace.
 \end{split}
 \end{equation*}
 
 In\cite{lempel_greenberg}, Lempel and Greenberg gave the first theoretical bound on $ H(\mathbf{X}) $, called the Lempel-Greenberg bound on an FHS.
 
 \begin{sec1_lemma1}[The Lempel-Greenberg bound \cite{lempel_greenberg}]\label{lempel_greenberg bound}
 For any FHS $ \mathbf{X} $ of length $ \nu $ over an alphabet of size $ m $, we have 
 \begin{equation*}
  H(\mathbf{X})\ge \big\lceil\dfrac{(\nu-b)(\nu+b-m)}{m(\nu-1)}\big\rceil
 \end{equation*}
 \end{sec1_lemma1}
 where $ b $ is the least nonnegative residue of $ \nu $ modulo $ m $, and $ \lceil r\rceil $ denotes the least integer no less than $ r $, a real number.
 
 The following result due to Fuji-Hara et al \cite{fujihara01} may be used to check the Lempel-Greenberg bound:
 \begin{sec1_cor1}[\cite{fujihara01}]
  For any FHS $ \mathbf{X} $ of length $ \nu $ over an alphabet of size $ m $,
  \begin{equation*}
   H(\mathbf{X})\ge
   \begin{cases}
   a,\quad \text{if}\ \nu\ne m\\
   0,\quad \text{if}\ \nu= m
   \end{cases}
  \end{equation*}
 \end{sec1_cor1}
where $ \nu=am+b,\ 0\le b<m $. 

\begin{sec1_def1}\label{LempelGreenbergOptimal}
An FHS $ \mathbf{X}\in \mathcal{S} $ is said to be optimal if $ \mathbf{X} $ is such that the equality in Lemma \ref{lempel_greenberg bound} is met.
\end{sec1_def1}
 
In \cite{peng_fan_bib1}, Peng and Fan established a bound on $  H(\mathcal{S}) $, resumed in the following lemma: 
\begin{sec1_lemma2}[The Peng-Fan bounds \cite{peng_fan_bib1}]\label{Peng-Fan-bounds}
Let $ \mathcal{S} $ be a set of $ M $ FHSs of length $ \nu $ over an alphabet of size $ m $, and $ I=\lfloor \nu M/m\rfloor $ where $ \lfloor r \rfloor $ denotes the integral part of $ r $. Then,
\begin{equation*}
H(\mathcal{S})\ge \bigg\lceil \dfrac{(\nu M-m)\nu}{(\nu M-1)m}\bigg\rceil
\end{equation*}
and
\begin{equation*}
H(\mathcal{S})\ge \bigg\lceil \dfrac{2I\nu M-(I+1)IM}{(\nu M-1)M}\bigg\rceil.
\end{equation*}
\end{sec1_lemma2}

\begin{sec1_def2}\label{PengFanOptimalDef}
The FHS set $ \mathcal{S} $ is said to be optimal if it meets one of the equalities of the Peng-Fan bounds in Lemma \ref{Peng-Fan-bounds}.
\end{sec1_def2}

Apart from the Hamming auto- and cross-correlation functions presented so far, the average Hamming correlation functions  are important as well to indicate the performance of the FHSs set, $ \mathcal{S} $. We at first define two overall numbers of the Hamming auto- and cross-correlation function as follows:
\begin{equation*}
\begin{split}
&\mathbf{N}_{a}( \mathcal{S})=\sum\limits_{\mathbf{X}\in  \mathcal{S}}\sum_{\tau=0}^{\nu-1}H(\mathbf{X}:\tau),\\
&\mathbf{N}_{c}( \mathcal{S})=\frac{1}{2}\sum\limits_{\mathbf{X},\mathbf{Y}\in  \mathcal{S},\mathbf{X}\ne \mathbf{Y}}\sum_{\tau=0}^{\nu-1}H(\mathbf{X},\mathbf{Y}:\tau)
\end{split}
\end{equation*}
From above two overall numbers can be defined the average Hamming auto- and  cross-correlation functions.
\begin{equation*}
\begin{split}
&\mathbf{A}_{a}( \mathcal{S})=\dfrac{\mathbf{N}_{a}( \mathcal{S})}{M(\nu-1)},\\
&\mathbf{A}_{c}( \mathcal{S})=\dfrac{2\mathbf{N}_{c}( \mathcal{S})}{\nu M(M-1)}.
\end{split}
\end{equation*}
 We recall $ M  $ is the number of the FHSs in the set $ \mathcal{S} $,  $ \nu $ is the length of each such FHS, and $ m $ is the size of the frequency alphabet set $ \mathcal{F} $. In a context not confused we write $ \mathbf{A}_{a} $ instead of $ \mathbf{A}_{a}( \mathcal{S}) $,  $ \mathbf{A}_{c} $ instead of $  \mathbf{A}_{c}( \mathcal{S}) $.  In \cite{PengPengAHP}, the authors gave a theoretical bound on $ \mathbf{A}_{a} $ and $ \mathbf{A}_{c} $ that relates other parameters $ M $, $ m $, and $ \nu $ together.
\begin{sec1_lemma3}[\cite{PengPengAHP}]\label{PPABound}
\begin{equation}\label{PPAB}
\dfrac{\mathbf{A}_{a}}{\nu (M-1)}+\dfrac{\mathbf{A}_{c}}{\nu -1}\geq \dfrac{\nu M-m}{m(\nu-1)(M-1)}
\end{equation}
\end{sec1_lemma3}

\begin{sec1_def3}\label{AHMOptimal}
The FHSs set $ \mathcal{S} $ is said to be optimal (AH Optimal) with respect to the average Hamming auto- and  cross-correlation functions if it is such that the equality in Lemma \ref{PPABound} is met.
\end{sec1_def3}

It is difficult and tedious to check if or not an FHS set is AH Optimal if we start up by computing explicitly $ \mathbf{A}_{a} $ and $ \mathbf{A}_{c} $ and then substitute them into (\ref{PPAB}). There is an indirect and efficient way  to verify the AH Optimality. We begin by introducing the concept of  an uniformly distributed FHS set \cite{ChungYangRef01,ChungYangRef02}.
\begin{sec1_def4}\label{lab_uniformly_distributed_FHS_set}
Let the symbols used here be that defined so far. The FHSs set $ \mathcal{S} $ is said to be an uniformly distributed FHSs set if $ \mathbf{N}_{ \mathcal{S}}(f) $ is a constant for any $ f\in \mathcal{F} $ where
\begin{equation*}
\mathbf{N}_{ \mathcal{S}}(f)=\sum\limits_{\mathbf{X} \in \mathcal{S}}\mathbf{N}_{\mathbf{X}}(f)
\end{equation*}
and
\begin{equation*}
\mathbf{N}_{\mathbf{X}}(f)=\arrowvert\lbrace 0\leq t\leq \nu-1:\mathbf{X}(t)=f \rbrace\arrowvert.
\end{equation*}
\end{sec1_def4}

Next theorem is the criterion to check if or not an FHSs set is AH Optimal.
\begin{sec1_thm1}[\cite{ChungYangRef01},\cite{ChungYangRef02}]\label{thm1_sec1}
The FHSs set $ \mathcal{S} $ is AH Optimal if only if it is  uniformly distributed.
\end{sec1_thm1}

Frequency-hopping sequences (FHSs) play an important role in communication systems such as frequency-hopping code-division multiple-access (FH-CDMA) systems, multi-user radar and sonar systems, etc.\cite{CDMA_FHSs}.  So, the construction of the FHSs with the optimal Hamming properties mentioned so far is an important research topics. There are several algebraic and combinatorial constructions in the literature \cite{ref_ding_01,ref_ding_02,ref_ding_03,ref_ding_04,fujihara01,fujihara02,fujihara03,tangxh01,tangxh02,tangxh03,tangxh04,tangxh05,tangxh06,tangxh07}. 

In this paper, we construct the FHSs of length $ p^{n} $ using  Ding-Helleseth generalized cyclotomy \cite{DingHellesethGCC}, show that the FHSs set is AH Optimal by the criterion stated in Theorem \ref{thm1_sec1}, and compute out explicitly the Hamming auto- and cross-correlation function. The rest of the paper is structured as follows: in Sect. 2, it is briefly introduced  Ding-Helleseth cyclotomy, based upon which the FHSs of length $ p^{n} $ are constructed, and their AH Optimal property is established. In Sect 3 and Sect. 4, it is given the formulae of the Hamming auto- and cross-correlation function of these FHSs. At the end of Sect. 4, we give an application to the case the length of the FHSs is equal to $ p^{3} $, and finally in Sect. 5,  some concluding remarks are presented.

\section{Ding-Helleseth Cyclotomy and Construction of the FHSs of Length $ p^{n} $}
Let $ n\geq 2 $ be an integer and $ Z_{n}^{*} $ be the set of all invertible elements of the additive group modulo $ n $, $ Z_{n} $. For any partition of $ Z_{n}^{*} $, $ Z_{n}^{*}=\bigcup_{i=0}^{d-1} D_{i}$ where $ D_{0} $ is a subgroup of $ Z_{n}^{*} $, if there exist $ d $ elements $ g_{1},g_{2},\cdots,g_{d-1} $, of $ Z_{n}^{*} $, such that $ D_{i}=g_{i}D_{0} $, then $ D_{i} $ is called a generalized cyclotomic class of order $ d $. In \cite{DingHellesethGCC}, Ding and Helleseth introduced a generalized cyclotomy with respect to $ n=p_{1}^{e_{1}}p_{2}^{e_{2}}\cdots p_{t}^{e_{t}} $, where $ p_{1},p_{2},\cdots,p_{t} $ are $ t $s distinct odd primes, and $ e_{1},e_{2},\cdots,e_{t} $ are $ t $s positive integers. Their initial aim was to extend  Whiteman generalized cyclotomy of $ pq $ \cite{whitemanGCC}, and construct balanced binary sequences for the use in cryptography.

Let $ p $ be an odd prime. It is known that if $ g $ is a primitive root modulo $ p^{2} $, then $ g $ is also a primitive root modulo $ p^{k} $, $ k\geq 1 $. By the Euler totient function, the order of $ g $ modulo $ p^{k} $ is equal to $ p^{k}-p^{k-1} $. Let $ D_{0}^{(p^{k})}=\langle g^{2}\rangle $ be the cyclic group generated by $ g^{2} $ modulo $ p^{k} $, and $ D_{1}^{(p^{k})}=g\langle g^{2}\rangle $ be the coset of $  D_{0}^{p^{k}} $ by $ g $. It is clear that both $ D_{0}^{(p^{k})} $ and $ D_{1}^{(p^{k})} $ are the Ding-Helleseth generalized cyclotomic class of order 2 which give a partition of the multiplicative group modulo $p^{k}$, $ Z_{p^{k}}^{*} $. The additive group modulo $ p^{n} $ can be decomposed into the union of  $ D_{0}^{(p^{k})} $s and $ D_{1}^{(p^{k})} $s, $ 1\leq k\leq n $ \cite[Lemma 12]{DingHellesethGCC}:
\begin{equation*}
Z_{p^{n}}\setminus\lbrace 0\rbrace=\biggl(\displaystyle\bigcup_{k=1}^{n}p^{n-k} D_{0}^{(p^{k})}\biggr)\cup\biggl(\displaystyle\bigcup_{k=1}^{n}p^{n-k} D_{1}^{(p^{k})}\biggr).
\end{equation*}

Define

\begin{equation}\label{DDCSet}
\begin{split}
&\mathcal{D}_{0}^{(k)}=p^{n-k}D_{0}^{(p^{k})},\ \mathcal{D}_{1}^{(k)}=p^{n-k}D_{1}^{(p^{k})},\\
&\mathcal{C}_{0}=\mathcal{D}_{0}^{(1)}\cup\left\lbrace 0\right\rbrace ,\ \mathcal{C}_{1}=\mathcal{D}_{1}^{(1)},\\
&\cdots\cdots\\
&\mathcal{C}_{2(k-1)}=\mathcal{D}_{0}^{(k)},\ \mathcal{C}_{2(k-1)+1}=\mathcal{D}_{1}^{(k)},\\
&\cdots\cdots\\
&\mathcal{C}_{2(n-1)}=\mathcal{D}_{0}^{(n)},\ \mathcal{C}_{2(n-1)+1}=\mathcal{D}_{1}^{(n)},\\
\end{split}
\end{equation}
where $ 1\leq k\leq n $. It is clear that with respect to $ Z_{p^{n}}^{*} $, there are $ m=2n $'s such sets: $ \mathcal{C}_{0},\mathcal{C}_{1},\cdots,\mathcal{C}_{2n-1} $.   $ \mathcal{C}_{i}\cap\mathcal{C}_{j}=\varPhi $ for $ 0\leq i\ne j\leq 2n-1 $ where $ \varPhi $ denotes the empty set, and $ Z_{p^{n}}=\bigcup_{i=0}^{2n-1}\mathcal{C}_{i} $.

Next, we describe the construction of the FHSs of length $ p^{n} $,  based on the Ding-Helleseth generalized cyclotomic classes of order 2 with respect to $ p^{n} $, where $ n\geq 3 $ and $ p $ is an odd prime.

Let $ \mathbf{X}=\bigl(\mathbf{X}(t)\bigr)_{t=0}^{\nu-1} $ be an FHS of length $ \nu $ over the frequency alphabet set $ \mathcal{F} $ of size $ m $. The support of $ f\in \mathcal{F} $ in the sequence $  \mathbf{X} $ is defined by
\begin{equation*}
\textbf{support}_{\mathbf{X}}(f)=\lbrace t|\mathbf{X}(t)=f,0\leq t\leq \nu-1 \rbrace.
\end{equation*}

\begin{sec2_cons1}\label{FHS_cons}
Let $ \mathcal{S}=\lbrace \mathbf{X}_{0},\mathbf{X}_{1},\cdots,\mathbf{X}_{2n-1}\rbrace $ be a set of FHSs of length $ p^{n} $. Let $ \mathbf{X}_{i},0\leq i\leq 2n-1 $, be such that 
\begin{equation*}
\textbf{support}_{\mathbf{X}_{i}}(j)=\mathcal{C}_{i+j}.
\end{equation*}
Where the subscript $ i+j $ is reduced modulo $ m=2n $.
\end{sec2_cons1}

It is obvious that Construct. \ref{FHS_cons} has the frequency alphabet set $ \mathcal{F}=\lbrace f|0\leq f\leq 2n-1\rbrace $, and the family size is equal to $ 2n $ as well. We have the following results:

\begin{sec2_thm1}\label{sec2_thm1_uniformed}
The FHSs Set, $  \mathcal{S} $, constructed from Construct. \ref{FHS_cons}, is  uniformed distributed.
\end{sec2_thm1}
\begin{proof}
Let $ j\in \mathcal{F}$. From Definition \ref{lab_uniformly_distributed_FHS_set} and Construct. \ref{FHS_cons}, 
\begin{equation*}
\begin{split}
\mathbf{N}_{ \mathcal{S}}(j)=\sum_{i=0}^{2n-1}\mathbf{N}_{\mathbf{X}_{i}}(j)
=\sum_{i=0}^{2n-1}|\mathcal{C}_{i+j}|=\sum_{k=0}^{2n-1}|\mathcal{C}_{k}|=p^{n}.
\end{split}
\end{equation*} 
So, for each  $ j\in \mathcal{F}$, $ \mathbf{N}_{ \mathcal{S}}(j) $ is constant. By Definition \ref{lab_uniformly_distributed_FHS_set}, the FHSs Set $  \mathcal{S} $  is  uniformed distributed.
\end{proof}

\begin{sec2_thm2}\label{sec2_thm2_AHOPT}
The FHSs Set, $  \mathcal{S} $, constructed from Construct. \ref{FHS_cons}, is AH Optimal.
\end{sec2_thm2} 
\begin{proof}
By Theorem \ref{thm1_sec1} and \ref{sec2_thm1_uniformed}.
\end{proof}

Define the generalized cyclotomic number of order two modulo $ p^{k} $  \cite{DingHellesethGCC} as below

\begin{equation*}
(i,j)_{p^{k}}=\biggl|\biggl(D^{(^{p^{k}})}_{i}+1\biggr)\cap D^{(^{p^{k}})}_{j}\biggr|
\end{equation*}
where $ i,j=0,1 $ and $ 1\leq k\leq n $. The formulae to compute above generalized cyclotomic numbers are given by the following equations \cite{DingHellesethGCC}:

If $ p\equiv 1\pmod 4 $, then
\begin{equation}\label{DH_cyclotomic_number_p1m4}
\begin{split}
(0,0)_{p^{k}}&=\dfrac{p^{k-1}(p-5)}{4},\\
(0,1)_{p^{k}}&=(1,0)_{p^{k}}=(1,1)_{p^{k}}=\dfrac{p^{k-1}(p-1)}{4}.
\end{split}
\end{equation}

If $ p\equiv 3\pmod 4 $, then
\begin{equation}\label{DH_cyclotomic_number_p1m5}
\begin{split}
(0,1)_{p^{k}}&=\dfrac{p^{k-1}(p-1)}{4},\\
(0,0)_{p^{k}}&=(1,0)_{p^{k}}=(1,1)_{p^{k}}=\dfrac{p^{k-1}(p-3)}{4}.
\end{split}
\end{equation}

In order to establish explicitly the Hamming auto-correlation function of the FHSs of length $ p^{n} $ in the sequel, we now define two types of distance functions:
\begin{equation}\label{dis_funcs}
\begin{split}
\Delta_{*,k}(i:\tau)&=\biggl|\lbrace 0\rbrace\cap \biggl(\mathcal{D}_{i}^{(k)}+\tau\biggr)\biggr|,\\
\Delta_{l,k}(i,j:\tau)&=\biggl| \mathcal{D}_{i}^{(l)}\cap \biggl(\mathcal{D}_{j}^{(k)}+\tau\biggr)\biggr|,
\end{split}
\end{equation}
where $ i,j=0,1 $,  $ 0\leq \tau < p^{n} $, and $ 1\leq l,k\leq n $.

We have the following lemmas related to $ \Delta_{*,k}(i:\tau) $ and $ \Delta_{l,k}(i,j:\tau) $:

\begin{sec2_lemma1}\label{sec2_lem1}
\begin{enumerate}
\item If $ p\equiv 1\pmod 4 $, then
\begin{equation*}
\Delta_{*,k}(i:\tau)=
\begin{cases}
1,&\ \text{if}\ \tau \in \mathcal{D}_{j}^{(k)}\ \text{and}\ j= i\\
0,&\ \text{otherwise}
\end{cases}
\end{equation*}
\item If $ p\equiv 3\pmod 4 $, then
\begin{equation*}
\Delta_{*,k}(i:\tau)=
\begin{cases}
1,&\ \text{if}\ \tau \in \mathcal{D}_{j}^{(k)}\ \text{and}\ j\ne i\\
0,&\ \text{otherwise}
\end{cases}
\end{equation*}
\end{enumerate}
\end{sec2_lemma1}
\begin{proof}
Proof for $ \Delta_{*,k}(0:\tau) $ is already given in \cite[Lemma 1]{JinPPowerN}. Proof for $ \Delta_{*,k}(1:\tau) $ is similar, so omitted.
\end{proof}

To compute  $ \Delta_{l,k}(i,j:\tau) $, three cases $ l<k,l=k $ and $ l>k $, are distinguished.

\begin{sec2_lemma2}\label{sec2_lam2}
\begin{enumerate}

\item $ l<k $. In this case, $ \Delta_{l,k}(0,0:\tau)= \Delta_{l,k}(1,0:\tau) $, and  $  \Delta_{l,k}(0,1:\tau)= \Delta_{l,k}(1,1:\tau) $.
\begin{enumerate}
\item $ p\equiv 1\pmod 4 $.
\begin{equation*}
 \Delta_{l,k}(0,0:\tau)= 
 \begin{cases}
 \frac{1}{2}(p^{l}-p^{l-1}),&\  \text{if}\ \tau \in \mathcal{D}_{0}^{(k)}\\
 0, &\  \text{otherwise}
 \end{cases}
\end{equation*}
\begin{equation*}
 \Delta_{l,k}(1,1:\tau)=
 \begin{cases}
 \frac{1}{2}(p^{l}-p^{l-1}),&\  \text{if}\ \tau \in \mathcal{D}_{1}^{(k)}\\
 0, &\  \text{otherwise}
 \end{cases}
\end{equation*}
\item $ p\equiv 3\pmod 4 $.
\begin{equation*}
 \Delta_{l,k}(0,0:\tau)= 
 \begin{cases}
 \frac{1}{2}(p^{l}-p^{l-1}),&\  \text{if}\ \tau \in \mathcal{D}_{1}^{(k)}\\
 0, &\  \text{otherwise}
 \end{cases}
\end{equation*}
\begin{equation*}
 \Delta_{l,k}(1,1:\tau)=
 \begin{cases}
 \frac{1}{2}(p^{l}-p^{l-1}),&\  \text{if}\ \tau \in \mathcal{D}_{0}^{(k)}\\
 0, &\  \text{otherwise}
 \end{cases}
\end{equation*}
\end{enumerate}

\item $ l=k $.
\begin{equation*}
 \Delta_{l,k}(0,0:\tau)= 
 \begin{cases}
 (0,0)_{p^{k}},&\  \text{if}\ \tau \in \mathcal{D}_{0}^{(k)}\\
 (1,1)_{p^{k}},&\  \text{if}\ \tau \in \mathcal{D}_{1}^{(k)}\\
 \frac{1}{2}(p^{k}-p^{k-1}),&\  \text{if}\ \tau \in \mathcal{D}_{0}^{(u)}\cup \mathcal{D}_{1}^{(u)}\\&\ \text{and}\ u<k\\
 0, &\  \text{otherwise}
 \end{cases}
\end{equation*}
\begin{equation*}
 \Delta_{l,k}(1,1:\tau)= 
 \begin{cases}
 (0,0)_{p^{k}},&\  \text{if}\ \tau \in \mathcal{D}_{1}^{(k)}\\
 (1,1)_{p^{k}},&\  \text{if}\ \tau \in \mathcal{D}_{0}^{(k)}\\
  \frac{1}{2}(p^{k}-p^{k-1}),&\  \text{if}\ \tau \in \mathcal{D}_{0}^{(u)}\cup \mathcal{D}_{1}^{(u)}\\&\ \text{and}\ u<k\\
 0, &\  \text{otherwise}
 \end{cases}
\end{equation*}
\begin{equation*}
 \Delta_{l,k}(1,0:\tau)= 
 \begin{cases}
 (0,1)_{p^{k}},&\  \text{if}\ \tau \in \mathcal{D}_{0}^{(k)}\\
 (1,0)_{p^{k}},&\  \text{if}\ \tau \in \mathcal{D}_{1}^{(k)}\\
 0, &\  \text{otherwise}
 \end{cases}
\end{equation*}
\begin{equation*}
 \Delta_{l,k}(0,1:\tau)= 
 \begin{cases}
 (1,0)_{p^{k}},&\  \text{if}\ \tau \in \mathcal{D}_{0}^{(k)}\\
 (0,1)_{p^{k}},&\  \text{if}\ \tau \in \mathcal{D}_{1}^{(k)}\\
 0, &\  \text{otherwise}
 \end{cases}
\end{equation*}
\item $ l>k $. In this case, $ \Delta_{l,k}(0,0:\tau)= \Delta_{l,k}(0,1:\tau) $,  $  \Delta_{l,k}(1,0:\tau)= \Delta_{l,k}(1,1:\tau) $, and
\begin{equation*}
 \Delta_{l,k}(0,0:\tau)= 
 \begin{cases}
 \frac{1}{2}(p^{k}-p^{k-1}),&\  \text{if}\ \tau \in \mathcal{D}_{0}^{(l)}\\
 0, &\  \text{otherwise}
 \end{cases}
\end{equation*}
\begin{equation*}
 \Delta_{l,k}(1,1:\tau)=
 \begin{cases}
 \frac{1}{2}(p^{k}-p^{k-1}),&\  \text{if}\ \tau \in \mathcal{D}_{1}^{(l)}\\
 0, &\  \text{otherwise}
 \end{cases}
\end{equation*}
\end{enumerate}
\end{sec2_lemma2}
\begin{proof}
Proof for $ \Delta_{l,k}(1,0:\tau) $ is already given in \cite[Lemma 2]{JinPPowerN}. Proof for other cases is  similar, so omitted.
\end{proof}

\section{Hamming Auto-correlation Function of the FHSs of Length $ p^{n} $}

\begin{sec3_thm1}\label{sec3_thm1_lab}
Let $ \mathbf{X}_{i}\in \mathcal{S} $ be an FHS generated by Construct. \ref{FHS_cons}, then its Hamming auto-correlation function can be determined according to two cases:
\begin{enumerate}
\item If $ p\equiv 1\pmod 4 $, then $ H(i:\tau)= $
\begin{equation*}\label{autocor_formula_p1m4}
\begin{cases}
\frac{1}{2}(2p^{n}-p+1),&\ \text{if}\ \tau\in \mathcal{D}_{0}^{(1)}\\
\frac{1}{2}(2p^{n}-p-3),&\ \text{if}\ \tau\in \mathcal{D}_{1}^{(1)}\\
\frac{1}{2}(2p^{n}-p^{k}-3p^{k-1}),&\ \text{if}\ \tau\in \mathcal{D}_{0}^{(k)}\cup\mathcal{D}_{1}^{(k)}\\&\ \text{and}\ 2\leq k\leq n
\end{cases}
\end{equation*}
\item If $ p\equiv 3\pmod 4 $, then $ H(i:\tau)= $
\begin{equation*}\label{autocor_formula_p3m4}
\begin{cases}
\frac{1}{2}(2p^{n}-p-1),&\ \text{if}\ \tau\in \mathcal{D}_{0}^{(1)}\cup\mathcal{D}_{1}^{(1)}\\
\frac{1}{2}(2p^{n}-p^{k}-3p^{k-1}),&\ \text{if}\ \tau\in \mathcal{D}_{0}^{(k)}\cup\mathcal{D}_{1}^{(k)}\\&\ \text{and}\ 2\leq k\leq n
\end{cases}
\end{equation*}
\end{enumerate}
\end{sec3_thm1}
\begin{proof}
It is clear that the number of the FHSs of length $ p^{n} $, constructed from Construct. \ref{FHS_cons}, is equal to $ 2n $. Let $ m=2n $, and $ \mathbf{X}_{i} $ be an FHS where $ 0\leq i\leq m-1 $. Then, the Hamming auto-correlation function of $ \mathbf{X}_{i} $, $ H(i:\tau) $ with $ 1\leq \tau<p^{n} $, can be computed as follows:
\begin{equation}\label{proof_sec3_thm1_00}
\begin{split}
H(i:\tau)&=\sum_{i=0}^{m-1}\left|\mathcal{C}_{i}\cap\left( \mathcal{C}_{i}+\tau\right)  \right|\\ 
&=\left|\mathcal{C}_{0}\cap\left( \mathcal{C}_{0}+\tau\right) \right| +\sum_{i=1}^{m-1}\left|\mathcal{C}_{i}\cap\left( \mathcal{C}_{i}+\tau\right)  \right|\\ 
&=\left|\mathcal{D}_{0}^{(1)}\cap\left( \mathcal{D}_{0}^{(1)}+\tau\right) \right| +\left| \mathcal{D}_{0}^{(1)}\cap\left\lbrace \tau\right\rbrace \right|\\
&\  +\left|\left\lbrace 0\right\rbrace \cap\left( \mathcal{D}_{0}^{(1)}+\tau\right) \right|+\sum_{i=1}^{m-1}\left|\mathcal{C}_{i}\cap\left( \mathcal{C}_{i}+\tau\right)  \right|\\ 
&=\left| \mathcal{D}_{0}^{(1)}\cap\left\lbrace \tau\right\rbrace \right|+\Delta_{*,1}(0:\tau)\\&\ +\sum_{k=1}^{n}\left|\mathcal{D}_{0}^{(k)}\cap\left( \mathcal{D}_{0}^{(k)}+\tau\right)  \right|\\&\ +\sum_{k=1}^{n}\left|\mathcal{D}_{1}^{(k)}\cap\left( \mathcal{D}_{1}^{(k)}+\tau\right)  \right|\\
&=\left| \mathcal{D}_{0}^{(1)}\cap\left\lbrace \tau\right\rbrace \right|+\Delta_{*,1}(0:\tau)\\&\ +\sum_{k=1}^{n}\Delta_{k,k}(0,0:\tau)+\sum_{k=1}^{n}\Delta_{k,k}(1,1:\tau)\\
\end{split}
\end{equation}
By the formulae for the case $ l=k $ in Lemma \ref{sec2_lam2}, it can be derived that 
\begin{equation}\label{proof_sec3_thm1_01}
\sum_{k=1}^{n}\Delta_{k,k}(0,0:\tau)=
\begin{cases}
(0,0)_{p^{k}}+\frac{1}{2}(p^{n}-p^{k}),&\ \text{if}\ \tau\in  \mathcal{D}_{0}^{(k)}\\
(1,1)_{p^{k}}+\frac{1}{2}(p^{n}-p^{k}),&\ \text{if}\ \tau\in  \mathcal{D}_{1}^{(k)}
\end{cases}
\end{equation}
and
\begin{equation}\label{proof_sec3_thm1_02}
\sum_{k=1}^{n}\Delta_{k,k}(1,1:\tau)=
\begin{cases}
(1,1)_{p^{k}}+\frac{1}{2}(p^{n}-p^{k}),&\ \text{if}\ \tau\in  \mathcal{D}_{0}^{(k)}\\
(0,0)_{p^{k}}+\frac{1}{2}(p^{n}-p^{k}),&\ \text{if}\ \tau\in  \mathcal{D}_{1}^{(k)}
\end{cases}
\end{equation}
Taking account of the value of  $ \Delta_{*,1}(0:\tau) $ from Lemma \ref{sec2_lem1} and substituting (\ref{proof_sec3_thm1_01})-(\ref{proof_sec3_thm1_02}) into the last equality of (\ref{proof_sec3_thm1_00}), we have
\begin{enumerate}
\item if $ p\equiv 1\pmod 4 $, then $ H(i:\tau)= $
\begin{equation}\label{proof_sec3_thm1_03}
\begin{cases}
2+(0,0)_{p}+(1,1)_{p}+p^{n}-p,&\ \text{if}\ \tau\in  \mathcal{D}_{0}^{(1)}\\
(0,0)_{p}+(1,1)_{p}+p^{n}-p,&\ \text{if}\ \tau\in  \mathcal{D}_{1}^{(1)}\\
(0,0)_{p^{k}}+(1,1)_{p^{k}}+p^{n}-p^{k},&\ \text{if}\ \tau\in  \mathcal{D}_{0}^{(k)}\cap\mathcal{D}_{1}^{(k)}\\ &\ \text{and}\ 2\leq k\leq n
\end{cases}
\end{equation}
\item if $ p\equiv 3\pmod 4 $, then $ H(i:\tau)= $
\begin{equation}\label{proof_sec3_thm1_04}
\begin{cases}
1+(0,0)_{p}+(1,1)_{p}+p^{n}-p,&\ \text{if}\ \tau\in  \mathcal{D}_{1}^{(0)}\cap\mathcal{D}_{1}^{(1)} \\
(0,0)_{p^{k}}+(1,1)_{p^{k}}+p^{n}-p^{k},&\ \text{if}\ \tau\in  \mathcal{D}_{0}^{(k)}\cap\mathcal{D}_{1}^{(k)}\\ &\ \text{and}\ 2\leq k\leq n
\end{cases}
\end{equation}
\end{enumerate}
Now, by substituting the formulae of the generalized cyclotomic numbers given in  (\ref{DH_cyclotomic_number_p1m4})-(\ref{DH_cyclotomic_number_p1m5}) into  (\ref{proof_sec3_thm1_03})-(\ref{proof_sec3_thm1_04}) respectively, we can establish the formulae in the actual Theorem. The proof is complete.
\end{proof}

\section{Hamming Cross-correlation Function of the FHSs of Length $ p^{n} $ for $ p\equiv 3\pmod 4 $}

Throughout this section, suppose that $ p $ is an odd prime and $ p\equiv 3\pmod 4 $.
Let $ \mathcal{S} $ be the FHS set constructed according to Construct. \ref{FHS_cons}, $ \mathbf{X}_{i}, \mathbf{X}_{j}\in \mathcal{S} $ be two distinct FHSs. Let $ \delta=(j-i)\pmod m $ where $ m=2n $,  $ \delta^{'}=\frac{\delta}{2} $ if $ \delta $ even, and $ \delta^{'}=\frac{\delta+1}{2} $ if $ \delta $ odd. Without loss of generality, suppose $ j>i $. The Hamming cross-correlation function can be determined according to various cases of $ \delta $: $ \delta $ even, $ \delta $ odd and $ \delta^{'}=1 $, $ \delta $ odd and $ \delta^{'}=n $, and $ \delta $ odd and $ 1<\delta^{'}<n $. 

\begin{sec3_prop1}\label{lab_sec3_prop01}
Suppose that $ \delta $ is odd and $ 1<\delta^{'}<n $. Then,
\begin{enumerate}
\item for $ n $ even and $ 2\delta^{'}=n $, $ H(i,i+\delta:\tau)= $
\begin{equation*}
\begin{cases}
0,&\ \text{if}\ \tau\in \lbrace 0\rbrace\\
0,&\ \text{if}\ \tau\in \mathcal{D}_{0}^{(k)}\cup \mathcal{D}_{1}^{(k)}\\&\ \text{and}\ 1\leq k<\delta^{'}\\
\frac{1}{2}(p+1), &\ \text{if}\ \tau\in \mathcal{D}_{0}^{(\delta^{'})}\\
0, &\ \text{if}\ \tau\in \mathcal{D}_{1}^{(\delta^{'})}\\
\frac{1}{2}p(p-1) &\ \text{if}\ \tau\in \mathcal{D}_{0}^{(\delta^{'}+1)}\\
p, &\ \text{if}\ \tau\in \mathcal{D}_{1}^{(\delta^{'}+1)}\\
\frac{1}{2}p^{k-\delta^{'}-2}(p^{2}+1)(p-1), &\ \text{if}\ \tau\in \mathcal{D}_{0}^{(k)}\\&\ \text{and}\ \delta^{'}+2\leq k\leq n\\
p^{k-\delta^{'}-1}(p-1), &\ \text{if}\ \tau\in \mathcal{D}_{1}^{(k)}\\&\ \text{and}\ \delta^{'}+2\leq k\leq n
\end{cases}
\end{equation*}
\item For $ n $ even and $ 2\delta^{'}=n+2 $, $ H(i,i+\delta:\tau)= $
\begin{equation*}
\begin{cases}
0,&\ \text{if}\ \tau\in \lbrace 0\rbrace\\
0,&\ \text{if}\ \tau\in \mathcal{D}_{0}^{(k)}\cup \mathcal{D}_{1}^{(k)}\\&\ \text{and}\ 1\leq k<\delta^{'}-1\\
0, &\ \text{if}\ \tau\in \mathcal{D}_{0}^{(\delta^{'}-1)}\\
\frac{1}{2}(p+1), &\ \text{if}\ \tau\in \mathcal{D}_{1}^{(\delta^{'}-1)}\\
p, &\ \text{if}\ \tau\in \mathcal{D}_{0}^{(\delta^{'})}\\
\frac{1}{2}p(p-1), &\ \text{if}\ \tau\in \mathcal{D}_{1}^{(\delta^{'})}\\
p^{k-\delta^{'}}(p-1), &\ \text{if}\ \tau\in \mathcal{D}_{0}^{(k)}\\&\ \text{and}\ \delta^{'}+1\leq k\leq n\\
\frac{1}{2}p^{k-\delta^{'}-1}(p^{2}+1)(p-1), &\ \text{if}\ \tau\in \mathcal{D}_{1}^{(k)}\\&\ \text{and}\ \delta^{'}+1\leq k\leq n
\end{cases}
\end{equation*}
\item For $ n $ odd and $ 2\delta^{'}=n+1 $, $ H(i,i+\delta:\tau)= $
\begin{equation*}
\begin{cases}
0,&\ \text{if}\ \tau\in \lbrace 0\rbrace\\
0,&\ \text{if}\ \tau\in \mathcal{D}_{0}^{(k)}\cup \mathcal{D}_{1}^{(k)} \\&\ \text{and}\ 1\leq k<\delta^{'}\\
\frac{1}{2}(p+1), &\ \text{if}\ \tau\in \mathcal{D}_{0}^{(\delta^{'})}\cup \mathcal{D}_{1}^{(\delta^{'})}\\
\frac{1}{2}p^{k-\delta^{'}-1}(p+1)(p-1), &\ \tau\in \mathcal{D}_{0}^{(k)}\cup\mathcal{D}_{1}^{(k)} \\&\ \text{and}\ \delta^{'}+1\leq k\leq n
\end{cases}
\end{equation*}
\item For  $ 2\delta^{'}<n $, let $ \epsilon=n-\delta^{'}+1 $. $ H(i,i+\delta:\tau)= $
\begin{equation*}
\begin{cases}
0,&\ \text{if}\ \tau\in \lbrace 0\rbrace\\
0,&\ \text{if}\ \tau\in \mathcal{D}_{0}^{(k)}\cup \mathcal{D}_{1}^{(k)}\\&\ \text{and}\ 1\leq k<\delta^{'}\\
\frac{1}{2}(p+1), &\ \text{if}\ \tau\in \mathcal{D}_{0}^{(\delta^{'})}\\
0, &\ \text{if}\ \tau\in  \mathcal{D}_{1}^{(\delta^{'})}\\
\frac{1}{2}p^{k-\delta^{'}}(p-1), &\ \text{if}\ \tau\in \mathcal{D}_{0}^{(k)}\ \text{and}\\&\ \delta^{'}+1\leq k< \epsilon+1\\
\frac{1}{2}p^{k-\epsilon-1}(p^{n+2-2\delta^{'}}+1)(p-1), &\ \text{if}\ \tau\in \mathcal{D}_{0}^{(k)}\\&\ \text{and}\ \epsilon+1\leq k\leq n\\
\frac{1}{2}p^{k-\delta^{'}-1}(p-1), &\ \text{if}\ \tau\in \mathcal{D}_{1}^{(k)} \ \text{and}\\&\ \delta^{'}+1\leq k< \epsilon\\
\frac{1}{2}(p^{n-2\delta^{'}+1}-p^{n-2\delta^{'}}+p+1),&\ \text{if}\ \tau\in \mathcal{D}_{1}^{(\epsilon)}\\
\frac{1}{2}p^{k-\epsilon}(p^{n-2\delta^{'}}+1)(p-1),&\ \text{if}\ \tau\in \mathcal{D}_{1}^{(k)}\\&\ \text{and}\ \epsilon< k\leq n
\end{cases}
\end{equation*}
\item For  $ 2\delta^{'}>n+2 $, let $ \epsilon=n-\delta^{'}+1 $. $ H(i,i+\delta:\tau)= $
\begin{equation*}
\begin{cases}
0,&\ \text{if}\ \tau\in \lbrace 0\rbrace\\
0,&\ \text{if}\ \tau\in \mathcal{D}_{0}^{(k)}\cup \mathcal{D}_{1}^{(k)}\\&\ \text{and}\ 1\leq k<\epsilon\\
0, &\ \text{if}\ \tau\in  \mathcal{D}_{0}^{(\epsilon)}\\
\frac{1}{2}(p+1), &\ \text{if}\ \tau\in \mathcal{D}_{1}^{(\epsilon)}\\
\frac{1}{2}p^{k-\epsilon-1}(p-1), &\ \text{if}\ \tau\in \mathcal{D}_{0}^{(k)}\ \text{and}\\&\ \epsilon+1\leq k<\delta^{'}\\
\frac{1}{2}(p^{\delta^{'}-\epsilon-1}(p-1)+p+1), &\ \text{if}\ \tau\in \mathcal{D}_{0}^{(\delta^{'})}\\
\frac{1}{2}p^{k-\epsilon-1}(p^{\epsilon-\delta^{'}+1}+1)(p-1), &\ \text{if}\ \tau\in \mathcal{D}_{0}^{(k)}\\&\ \text{and}\ \delta^{'}< k\leq n\\
\frac{1}{2}p^{k-\epsilon}(p-1), &\ \text{if}\ \tau\in \mathcal{D}_{1}^{(k)} \ \text{and}\\&\ \epsilon< k< \delta^{'}+1\\
\frac{1}{2}p^{k-\epsilon}(p^{\epsilon-\delta^{'}}+1)(p-1),&\ \text{if}\ \tau\in \mathcal{D}_{1}^{(k)}\\&\ \text{and}\ \delta^{'}+1\leq k\leq n
\end{cases}
\end{equation*}
\end{enumerate}
\end{sec3_prop1}
\begin{proof}
From Construct. \ref{FHS_cons} and (\ref{DDCSet}), it is clear that
\begin{equation}\label{Sec3Propo2Proof01}
\begin{split}
H(i,i+\delta:\tau)&=\sum_{k=0}^{m-1}\left| \mathcal{C}_{k}\cap(\mathcal{C}_{k+\delta}+\tau)\right|\\
&= \sum_{k=1}^{n}\left| \mathcal{C}_{2(k-1)}\cap(\mathcal{C}_{2(k-1)+\delta}+\tau)\right|\\
&\ +\sum_{k=1}^{n}\left| \mathcal{C}_{2(k-1)+1}\cap(\mathcal{C}_{2(k-1)+1+\delta}+\tau)\right|
\end{split}
\end{equation}
In (\ref{Sec3Propo2Proof01}), $ \mathcal{C}_{0} $ occurs two times at $ k=1 $ and $ k=(m-(\delta+1))/2+1=n-\delta^{'}+1 $, respectively. The corresponding items in the summation are $ \left| \mathcal{C}_{0}\cap(\mathcal{C}_{\delta}+\tau)\right|=|(\mathcal{D}_{0}^{(1)} \cup \lbrace 0\rbrace)\cap(\mathcal{D}_{1}^{(\delta^{'})}+\tau)|   $ and $ | \mathcal{C}_{m-\delta}\cap(\mathcal{C}_{0}+\tau)|=|\mathcal{D}_{1}^{(n-\delta^{'}+1)} \cap((\mathcal{D}_{0}^{(1)}+\tau)\cup \lbrace\tau\rbrace)| $, respectively. From above analysis and (\ref{Sec3Propo2Proof01}), we pursue
\begin{equation}\label{Sec3Propo2Proof02}
\begin{split}
& H(i,i+\delta:\tau)\\
 &= \sum_{k=1}^{n}| \mathcal{D}_{0}^{(k)}\cap(\mathcal{D}_{1}^{(k+\delta^{'}-1)}+\tau)|
+| \lbrace 0\rbrace \cap (\mathcal{D}_{1}^{(\delta^{'})}+\tau)|\\
&\ + \sum_{k=1}^{n}| \mathcal{D}_{1}^{(k)}\cap(\mathcal{D}_{0}^{(k+\delta^{'})}+\tau)|
+| \mathcal{D}_{1}^{(n-\delta^{'}+1)}\cap \lbrace \tau\rbrace |\\
&=\sum_{k=1}^{n}\Delta_{k,k+\delta^{'}-1}(0,1:\tau)+\Delta_{*,\delta^{'}}(1:\tau)\\
&\ + \sum_{k=1}^{n}\Delta_{k,k+\delta^{'}}(1,0:\tau)+| \mathcal{D}_{1}^{(n-\delta^{'}+1)}\cap \lbrace \tau\rbrace |
\end{split}
\end{equation}
In the last equation of (\ref{Sec3Propo2Proof02}), each summation can be split into two parts whose generic items, $ \Delta_{l,k}(i,j:\tau) $, correspond to cases $ l<k $ and $ l>k $, respectively. From above analysis and (\ref{Sec3Propo2Proof02}), we have
\begin{equation}\label{Sec3Propo2Proof03}
\begin{split}
& H(i,i+\delta:\tau)\\
&=\sum_{k=1}^{n-\delta^{'}+1}\Delta_{k,k+\delta^{'}-1}(0,1:\tau)+\sum_{k=1}^{\delta^{'}-1}\Delta_{n-\delta^{'}+k+1,k}(0,1:\tau)\\
&\ + \sum_{k=1}^{n-\delta^{'}}\Delta_{k,k+\delta^{'}}(1,0:\tau)+\sum_{k=1}^{\delta^{'}}\Delta_{n-\delta^{'}+k,k}(1,0:\tau)\\
&\ +\Delta_{*,\delta^{'}}(1:\tau)+|\mathcal{D}_{1}^{(n-\delta^{'}+1)}\cap \lbrace \tau\rbrace |\\
\end{split}
\end{equation}
By Lemma \ref{sec2_lam2}, each summation in (\ref{Sec3Propo2Proof03}) can be explicitly written down:

$ \sum_{k=1}^{n-\delta^{'}+1}\Delta_{k,k+\delta^{'}-1}(0,1:\tau)= $
\begin{equation}\label{Sec3Propo2Proof031}
\begin{cases}
\frac{1}{2}(p^{k-\delta^{'}+1}-p^{k-\delta^{'}}),&\ \text{if}\ \tau\in \mathcal{D}_{0}^{(k)}\\&\ \text{and}\ \delta^{'}\leq k\leq n\\
0,&\ \text{otherwise}
\end{cases}
\end{equation}

$ \sum_{k=1}^{\delta^{'}-1}\Delta_{n-\delta^{'}+k+1,k}(0,1:\tau)= $
\begin{equation}\label{Sec3Propo2Proof032}
\begin{cases}
\frac{1}{2}(p^{k+\delta^{'}-n-1}-p^{k+\delta^{'}-n-2}),&\ \text{if}\ \tau\in \mathcal{D}_{0}^{(k)}\\&\ \text{and}\ n-\delta^{'}+2\leq k\leq n\\
0,&\ \text{otherwise}
\end{cases}
\end{equation}

$ \sum_{k=1}^{n-\delta^{'}}\Delta_{k,k+\delta^{'}}(1,0:\tau)= $
\begin{equation}\label{Sec3Propo2Proof033}
\begin{cases}
\frac{1}{2}(p^{k-\delta^{'}}-p^{k-\delta^{'}-1}),&\ \text{if}\ \tau\in \mathcal{D}_{1}^{(k)}\\&\ \text{and}\ \delta^{'}+1\leq k\leq n\\
0,&\ \text{otherwise}
\end{cases}
\end{equation}

$\sum_{k=1}^{\delta^{'}}\Delta_{n-\delta^{'}+k,k}(1,0:\tau)= $
\begin{equation}\label{Sec3Propo2Proof034}
\begin{cases}
\frac{1}{2}(p^{k+\delta^{'}-n}-p^{k+\delta^{'}-n-1}),&\ \text{if}\ \tau\in \mathcal{D}_{1}^{(k)}\\&\ \text{and}\ n-\delta^{'}+1\leq k\leq n\\
0,&\ \text{otherwise}
\end{cases}
\end{equation}
Consider the lower bounds of $ k $'s in (\ref{Sec3Propo2Proof031})-(\ref{Sec3Propo2Proof034}). It leads to five possibilities: $ n $ even and $ 2\delta^{'}=n $,  which indicates the lower bound of (\ref{Sec3Propo2Proof033}) meets the one of (\ref{Sec3Propo2Proof034}); $ n $ even and $ 2\delta^{'}=n+2 $, which indicates the lower bound of (\ref{Sec3Propo2Proof031}) meets the one of (\ref{Sec3Propo2Proof032}); $ n $ odd and $ 2\delta^{'}=n+1 $, which indicates the lower bound of (\ref{Sec3Propo2Proof031}) meets the one of (\ref{Sec3Propo2Proof034}); $ 2\delta^{'}<n $, and $ 2\delta^{'}>n+2 $. Further analysis of those cases together with help of Lemma \ref{sec2_lem1} is straightforward, so omitted.
\end{proof}

\begin{sec3_prop2}\label{lab_sec3_prop2}
Suppose that $ \delta $ is odd. Then,
\begin{enumerate}
\item for $ \delta^{'}=1 $, $ H(i,i+1:\tau)= $
\begin{equation*}
\begin{cases}
0&\ \text{if}\ \tau=0,\\
\frac{1}{4}(p+1), &\ \text{if}\ \tau\in \mathcal{D}_{0}^{(1)}\cup \mathcal{D}_{1}^{(1)}\\
\frac{1}{4}p^{k-1}(p-3), &\ \text{if}\ \tau\in \mathcal{D}_{0}^{(k)}\ \text{and}\\&\ 2\leq k\leq n\\
\frac{1}{4}p^{k-2}(p^{2}+3p-2), &\ \text{if}\ \tau\in \mathcal{D}_{1}^{(k)}\ \text{and}\\&\ 2\leq k< n\\
\frac{1}{4}(p^{n-2}(p^{2}+3p-2)+2p+2), &\ \text{if}\ \tau\in \mathcal{D}_{1}^{(n)}
\end{cases}
\end{equation*}
\item For $ \delta^{'}=n $, $ H(i,i+2n-1:\tau)= $
\begin{equation*}
\begin{cases}
0,&\ \text{if}\ \tau=0\\
\frac{1}{4}(p+1), &\ \text{if}\ \tau\in \mathcal{D}_{0}^{(1)}\cup \mathcal{D}_{1}^{(1)}\\
\frac{1}{4}p^{k-2}(p^{2}+3p-2), &\ \text{if}\ \tau\in \mathcal{D}_{0}^{(k)}\ \text{and}\\&\ 2\leq k< n\\
\frac{1}{4}p^{k-1}(p-3), &\ \text{if}\ \tau\in \mathcal{D}_{1}^{(k)}\ \text{and}\\&\ 2\leq k< n\\
\frac{1}{4}(p^{n-2}(p^{2}+3p-2)+2p+2), &\ \text{if}\ \tau\in \mathcal{D}_{0}^{(n)}\\
\frac{1}{4}p^{n-1}(p-3),&\ \text{if}\ \tau\in \mathcal{D}_{1}^{(n)}
\end{cases}
\end{equation*}
\end{enumerate}
\end{sec3_prop2}
\begin{proof}
From Construct. \ref{FHS_cons} and (\ref{DDCSet}), we can obtain 
\begin{equation*}
\begin{split}
&\ H(i,i+1:\tau)\\
&=\Delta_{*,1}(1:\tau)+|\mathcal{D}_{1}^{(n)}\cap\lbrace \tau\rbrace | \\
&\ +\sum_{k=1}^{n}\Delta_{k,k}(0,1:\tau)+\sum_{k=1}^{n}\Delta_{k,k+1}(1,0:\tau),\\
&\ H(i,i+2n-1:\tau)\\
&=\Delta_{*,n}(1:\tau)+|\mathcal{D}_{1}^{(1)}\cap\lbrace \tau\rbrace | \\
&\ +\sum_{k=1}^{n}\Delta_{k,k}(1,0:\tau)+\sum_{k=1}^{n}\Delta_{k,k-1}(0,1:\tau).
\end{split}
\end{equation*}
Further analysis is similar to the proof of Proposition \ref{lab_sec3_prop01}, so omitted.
\end{proof}

\begin{sec3_prop3}\label{lab_sec3_prop03}
Suppose that $ \delta $ is even. Then,
\begin{enumerate}
\item for $ n $ even and $ 2\delta^{'}=n $, $ H(i,i+\delta:\tau)= $
\begin{equation*}
\begin{cases}
0,&\ \text{if}\ \tau\in \lbrace 0\rbrace\\
0,&\ \text{if}\ \tau\in \mathcal{D}_{0}^{(k)}\cup \mathcal{D}_{1}^{(k)}\ \text{and}\\&\ 1\leq k<\delta^{'}+1\\
p, &\ \text{if}\ \tau\in \mathcal{D}_{0}^{(\delta^{'}+1)}\cup\mathcal{D}_{1}^{(\delta^{'}+1)} \\
p^{k-\delta^{'}-1}(p-1), &\ \text{if}\ \tau\in \mathcal{D}_{0}^{(k)}\cup\mathcal{D}_{1}^{(k)}\ \text{and}\\&\ \delta^{'}+1< k\leq n
\end{cases}
\end{equation*}
\item For  $ 2\delta^{'}<n $, let $ \epsilon=n-\delta^{'}+1 $. Then, $ H(i,i+\delta:\tau)= $
\begin{equation*}
\begin{cases}
0,&\ \text{if}\ \tau\in \lbrace 0\rbrace\\
0,&\ \text{if}\ \tau\in \mathcal{D}_{0}^{(k)}\cup \mathcal{D}_{1}^{(k)}\ \text{and}\\&\ 1\leq k<\delta^{'}+1\\
\frac{1}{2}(p-1),&\ \text{if}\ \tau\in \mathcal{D}_{0}^{(\delta^{'}+1)}\\
\frac{1}{2}(p+1),&\ \text{if}\ \tau\in \mathcal{D}_{1}^{(\delta^{'}+1)}\\
\frac{1}{2}p^{k-\delta^{'}-1}(p-1), &\ \text{if}\ \tau\in \mathcal{D}_{0}^{(k)}\cup\mathcal{D}_{1}^{(k)}\ \text{and}\\&\ \delta^{'}+1< k< \epsilon\\
\frac{1}{2}(p^{n-2\delta^{'}}(p-1)+p+1),&\ \text{if}\ \tau\in \mathcal{D}_{0}^{(\epsilon)}\\
\frac{1}{2}(p^{n-2\delta^{'}}+1)(p-1),&\ \text{if}\ \tau\in \mathcal{D}_{1}^{(\epsilon)}\\
\frac{1}{2}p^{k-\epsilon}(p^{n-2\delta^{'}}+1)(p-1), &\ \text{if}\ \tau\in \mathcal{D}_{0}^{(k)}\cup\mathcal{D}_{0}^{(k)}\ \text{and}\\&\ \epsilon< k\leq n\\
\end{cases}
\end{equation*}
\item For  $ 2\delta^{'}>n $, let $ \epsilon=n-\delta^{'}+1 $.  $ H(i,i+\delta:\tau)= $
\begin{equation*}
\begin{cases}
0,&\ \text{if}\ \tau\in \lbrace 0\rbrace\\
0,&\ \text{if}\ \tau\in \mathcal{D}_{0}^{(k)}\cup \mathcal{D}_{1}^{(k)}\\&\ \text{and}\  1\leq k<\epsilon\\
\frac{1}{2}(p+1),&\ \text{if}\ \tau\in \mathcal{D}_{0}^{(\epsilon)}\\
\frac{1}{2}(p-1),&\ \text{if}\ \tau\in \mathcal{D}_{1}^{(\epsilon)}\\
\frac{1}{2}p^{k-\epsilon}(p-1), &\ \text{if}\ \tau\in \mathcal{D}_{0}^{(k)}\cup\mathcal{D}_{1}^{(k)}\ \text{and}\\&\ \epsilon< k<\delta^{'}+1\\
\frac{1}{2}(p^{2\delta^{'}-n}+1)(p-1),&\ \text{if}\ \tau\in \mathcal{D}_{0}^{(\delta^{'}+1)}\\
\frac{1}{2}(p^{2\delta^{'}-n}(p-1)+p+1),&\ \text{if}\ \tau\in \mathcal{D}_{1}^{(\delta^{'}+1)}\\
\frac{1}{2}p^{k-\epsilon}(p^{n-2\delta^{'}}+1)(p-1), &\ \text{if}\ \tau\in \mathcal{D}_{0}^{(k)}\cup\mathcal{D}_{1}^{(k)}\ \text{and}\\&\ \delta^{'}+1< k\leq n
\end{cases}
\end{equation*}
\end{enumerate}
\end{sec3_prop3}
\begin{proof}
From Construct. \ref{FHS_cons} and (\ref{DDCSet}), we can obtain 
\begin{equation*}
\begin{split}
&\ H(i,i+\delta:\tau)\\
&=\Delta_{*,\delta^{'}+1}(0:\tau)+|\mathcal{D}_{0}^{(n-\delta^{'}+1)}\cap\lbrace \tau\rbrace | \\
&\ +\sum_{k=1}^{n}\Delta_{k,k+\delta^{'}}(0,0:\tau)+\sum_{k=1}^{n}\Delta_{k,k+\delta^{'}}(1,1:\tau).
\end{split}
\end{equation*}
Further analysis is similar to the proof of Proposition \ref{lab_sec3_prop01}, so omitted.
\end{proof}

The Hamming cross-correlation function of six FHSs of length $ p^{3} $ can be derived from Construct. \ref{FHS_cons} and Proposition \ref{lab_sec3_prop01}-\ref{lab_sec3_prop03}:

\begin{sec3_cor1}\label{lab_sec3_cor1}
Let $ \mathbf{X}_{i} $, $ \mathbf{X}_{j} $ be two distinct FHSs of length $ p^{3} $ constructed according to Construct. \ref{FHS_cons}, then the Hamming cross-correlation function between  $ \mathbf{X}_{i} $ and $ \mathbf{X}_{j} $, $ H(i,j:\tau)=H(i,i+(j-i)\pmod 6:\tau) $, is given by the following equations:
\begin{enumerate}
\item Let $ \delta=(j-i)\pmod 6=1 $, then $ H(i,i+1:\tau)= $
\begin{equation*}
\begin{cases}
0,&\ \text{if}\ \tau=0\\
\frac{1}{4}(p+1),&\ \text{if}\ \tau\in \mathcal{D}_{0}^{(1)}\cup \mathcal{D}_{1}^{(1)}\\
\frac{1}{4}p(p-3),&\ \text{if}\ \tau\in \mathcal{D}_{0}^{(2)}\\
\frac{1}{4}(p^{2}+3p-2),&\ \text{if}\ \tau\in \mathcal{D}_{1}^{(2)}\\
\frac{1}{4}p^{2}(p-3),&\ \text{if}\ \tau\in \mathcal{D}_{0}^{(3)}\\
\frac{1}{4}(p^{3}+3p^{2}+2),&\ \text{if}\ \tau\in \mathcal{D}_{1}^{(3)}
\end{cases}
\end{equation*}
\item Let $ \delta=(j-i)\pmod 6=5 $, then $ H(i,i+5:\tau)= $
\begin{equation*}
\begin{cases}
0,&\ \text{if}\ \tau=0\\
\frac{1}{4}(p+1),&\ \text{if}\ \tau\in \mathcal{D}_{0}^{(1)}\cup \mathcal{D}_{1}^{(1)}\\
\frac{1}{4}(p^{2}+3p-2),&\ \text{if}\ \tau\in \mathcal{D}_{0}^{(2)}\\
\frac{1}{4}p(p-3),&\ \text{if}\ \tau\in \mathcal{D}_{1}^{(2)}\\
\frac{1}{4}(p^{3}+3p^{2}+2),&\ \text{if}\ \tau\in \mathcal{D}_{0}^{(3)}\\
\frac{1}{4}p^{2}(p-3),&\ \text{if}\ \tau\in \mathcal{D}_{1}^{(3)}
\end{cases}
\end{equation*}
\item Let $ \delta=(j-i)\pmod 6=3 $, then $ H(i,i+3:\tau)= $
\begin{equation*}
\begin{cases}
0,&\ \text{if}\ \tau=0\\
0,&\ \text{if}\ \tau\in \mathcal{D}_{0}^{(1)}\cup \mathcal{D}_{1}^{(1)}\\
\frac{1}{2}(p+1),&\ \text{if}\ \tau\in \mathcal{D}_{0}^{(2)}\cup \mathcal{D}_{1}^{(2)}\\
\frac{1}{2}(p+1)(p-1),&\ \text{if}\ \tau\in \mathcal{D}_{0}^{(3)}\cup \mathcal{D}_{1}^{(3)}
\end{cases}
\end{equation*}
\item Let $ \delta=(j-i)\pmod 6=2 $, then $ H(i,i+2:\tau)= $
\begin{equation*}
\begin{cases}
0,&\ \text{if}\ \tau=0\\
0,&\ \text{if}\ \tau\in \mathcal{D}_{0}^{(1)}\cup \mathcal{D}_{1}^{(1)}\\
\frac{1}{2}(p-1),&\ \text{if}\ \tau\in \mathcal{D}_{0}^{(2)}\\
\frac{1}{2}(p+1),&\ \text{if}\ \tau\in \mathcal{D}_{1}^{(2)}\\
\frac{1}{2}(p^{2}+1),&\ \text{if}\ \tau\in \mathcal{D}_{0}^{(3)}\\
\frac{1}{2}(p+1)(p-1),&\ \text{if}\ \tau\in \mathcal{D}_{1}^{(3)}
\end{cases}
\end{equation*}
\item Let $ \delta=(j-i)\pmod 6=4 $, then $ H(i,i+4:\tau)= $
\begin{equation*}
\begin{cases}
0,&\ \text{if}\ \tau=0\\
0,&\ \text{if}\ \tau\in \mathcal{D}_{0}^{(1)}\cup \mathcal{D}_{1}^{(1)}\\
\frac{1}{2}(p+1),&\ \text{if}\ \tau\in \mathcal{D}_{0}^{(2)}\\
\frac{1}{2}(p-1),&\ \text{if}\ \tau\in \mathcal{D}_{1}^{(2)}\\
\frac{1}{2}(p+1)(p-1),&\ \text{if}\ \tau\in \mathcal{D}_{0}^{(3)}\\
\frac{1}{2}(p^{2}+1),&\ \text{if}\ \tau\in \mathcal{D}_{1}^{(3)}
\end{cases}
\end{equation*}
\end{enumerate}
\end{sec3_cor1}

\begin{proof}
For each case of $ \delta=(j-i)\pmod 6 $, substitute $ n=3 $ in the corresponding Proposition from \ref{lab_sec3_prop01}-\ref{lab_sec3_prop03}, and take care of the interval of $ \tau $, that ranges from $ 0 $ to $ p^{3}-1 $.
\end{proof}

\begin{sec4_thm1}\label{lab-sec4-thm1}
Let $ \mathbf{X}_{i}, \mathbf{X}_{j}\in \mathcal{S} $ be two FHSs generated by Construct. \ref{FHS_cons} with $ i\ne j $. Then, their Hamming cross-correlation function is uniquely determined by the formulas given by Proposition \ref{lab_sec3_prop01}-\ref{lab_sec3_prop03}.
\end{sec4_thm1}
\begin{proof}
Let $ \mathcal{S} $ be the FHS set constructed according to Construct. \ref{FHS_cons}, $ \mathbf{X}_{i}, \mathbf{X}_{j}\in \mathcal{S} $ be two distinct FHSs. Let $ \delta=(j-i)\pmod m $ where $ m=2n $,  $ \delta^{'}=\frac{\delta}{2} $ if $ \delta $ even, and $ \delta^{'}=\frac{\delta+1}{2} $ if $ \delta $ odd. It is clear that Proposition \ref{lab_sec3_prop01}-\ref{lab_sec3_prop03} cover all the possible cases  $ \delta $ may take.
\end{proof}
\section{Conclusion}

In this paper, a new class of the frequency-hopping sequences (FHSs) of length $ p^{n} $ is constructed based on Ding-Helleseth generalized cyclotomic classes of order two, of which the Hamming auto- and cross-correlation functions are established (for the Hamming cross-correlation, only the case $ p\equiv 3\pmod 4 $ is considered). It is shown that the constructed FHSs' set is uniformly distributed, and optimal with respect to the average Hamming correlation functions.


\begin{thebibliography}{99}
\bibitem{lempel_greenberg}
A. Lempel, and H. Greenberg, \textquotedblleft Families of sequences with optimal Hamming correlation properties,\textquotedblright $ \  $IEEE Trans. Inf. Theory, vol.20, no.1, pp.90-94, 1974.
\bibitem{peng_fan_bib1}
D. Y. Peng, and P. Z. Fan, \textquotedblleft Lower bounds on the Hamming auto- and cross correlations of frequency-hopping sequences,\textquotedblright $ \  $IEEE Trans. Inf. Theory, vol.50, no.9, pp.2149-21544, Sept. 2004.
\bibitem{fujihara01}
R. Fuji-Hara, Y. Miao, and M. Mishima, \textquotedblleft Optimal frequency hopping sequences: A combinatorial approach,\textquotedblright $ \  $IEEE Trans. Inf. Theory, vol.50, no.10, pp.2408-2420, Oct. 2004.
\bibitem{PengPengAHP}
D. Peng, T. Peng, X. Tang, and X. Niu, \textquotedblleft Aclass of optimal frequency hopping sequences based on the theory of power residues,\textquotedblright $ \  $Lect. Notes Comput. Sci., Seque. and their applic., vol.5203,  pp.188-196, Oct. 2008.
\bibitem{ChungYangRef01}
J. H. Chung, and K. Yang, \textquotedblleft New frequency-hopping sequence sets with optimal average and good maximum Hamming correlations,\textquotedblright $ \  $IET Communications, vol.6, no.13,  pp.2048-2053, Sep. 2012.
\bibitem{ChungYangRef02}
J. H. Chung, and K. Yang, \textquotedblleft Frequency-hopping sequence sets with low average and  maximum Hamming correlations,\textquotedblright $ \  $\textit{arXiv:1108.3415}.
\bibitem{CDMA_FHSs}
M. K. Simon, J. K. Omura, R. A. Schotz, and B. K. Levitt, Spread Spectrum Communications, vol.1, Rockville MD: Computer Science Press, 1985.
\bibitem{ref_ding_01}
C. Ding, M. J. Moisio, and J. Yuan, \textquotedblleft Algebraic constructions of optimal frequency-hopping sequences,\textquotedblright $ \  $IEEE Trans. Inf. Theory, vol.53, no.7, pp.2606-2610, Jul. 2007.
\bibitem{ref_ding_02}
C. Ding, and J. Yuan, \textquotedblleft Sets of optimal frequency-hopping sequences,\textquotedblright $ \  $IEEE Trans. Inf. Theory, vol.54, no.8, pp.3741-3745, Aug. 2008.
\bibitem{ref_ding_03}
C. Ding, R. Fuji-Hara, Y. Fujiwaha, M. Jimbo, and M. Mishima, \textquotedblleft Sets of  frequency-hopping sequences: Bounds and optimal constructions,\textquotedblright $ \  $IEEE Trans. Inf. Theory, vol.55, no.7, pp.3297-3304, Jul. 2009.
\bibitem{ref_ding_04}
C. Ding, Y. Yang, and X. Tang,  \textquotedblleft Optimal Sets of Frequency Hopping Sequences From Linear Cyclic Codes,\textquotedblright $ \  $IEEE Trans. Inf. Theory,  vol.56, no.7, pp.3605-3612, July 2010.
\bibitem{fujihara02}
G. Ge, R. Fuji-Hara, and Y. Miao, \textquotedblleft Further combinatorial constructions for optimal frequency hopping sequences,\textquotedblright $ \  $J. Combin. Theory Ser. A, vol.113, pp.1699-1718,  2006.
\bibitem{fujihara03}
G. Ge, Y. Miao, and Z. H. Yao,  \textquotedblleft Optimal frequency hopping sequences: Auto- and cross-correlation properties,\textquotedblright $ \  $IEEE Trans. Inf. Theory,  vol.55, no.2, pp.867-879, Feb. 2009.
\bibitem{tangxh01}
Z. Zhou, X. Tang, D. Peng, and U. Parampalli,  \textquotedblleft New Constructions for Optimal Sets of Frequency-Hopping Sequences,\textquotedblright $ \  $IEEE Trans. Inf. Theory,  vol.57, no.6, pp.3831-3840, June 2011.
\bibitem{tangxh02}
X. Zeng, H. Cai, X. Tang, and Y. Yang,  \textquotedblleft Optimal Frequency Hopping Sequences of Odd Length,\textquotedblright $ \  $IEEE Trans. Inf. Theory,  vol.59, no.5, pp.3237-3248, May 2013.
\bibitem{tangxh03}
J. H. Chung, G. Gong, and K. Yang,  \textquotedblleft New Families of Optimal Frequency-Hopping Sequences of Composite Lengths,\textquotedblright $ \  $IEEE Trans. Inf. Theory,  vol.60, no.6, pp.3688-3697, June 2014.
\bibitem{tangxh04}
H. Y. Han, and D. Y. Peng,  \textquotedblleft Set of optimal frequency-hopping sequences based on polynomial theory,\textquotedblright $ \  $Electronics Letters, vol.50, no.3, pp.214-216, Jan. 2014.
\bibitem{tangxh05}
H. Cai, Z. Zhou, Y. Yang, and X. Tang,  \textquotedblleft A New Construction of Frequency-Hopping Sequences With Optimal Partial Hamming Correlation,\textquotedblright $ \  $IEEE Trans. Inf. Theory, vol.60, no.9, pp.5782-5790, Sept. 2014.
\bibitem{tangxh06}
C. Fan, H. Cai, and X. Tang,  \textquotedblleft A Combinatorial Construction for Strictly Optimal Frequency-Hopping Sequences,\textquotedblright $ \  $IEEE Trans. Inf. Theory, vol.62, no.8, pp.4769-4774, Aug. 2016.
\bibitem{tangxh07}
J. Bao, and L. Ji,  \textquotedblleft New Families of Optimal Frequency Hopping Sequence Sets,\textquotedblright $ \  $IEEE Trans. Inf. Theory, vol.62, no.9, pp.5209-5224, Sept. 2016.
\bibitem{whitemanGCC}
A. L. Whiteman,  \textquotedblleft A family of difference sets,\textquotedblright $ \  $Illinois J. Math, vol.6,  pp.107-21, 1962. 
\bibitem{DingHellesethGCC}
C. Ding, and T. Helleseth,  \textquotedblleft New generalized cyclotomy and its applications,\textquotedblright $ \  $Finite Fields and Their Applications, vol.4,  pp.140-166, 1998. 
\bibitem{JinPPowerN}
S. Y. Jin, Y. J. Kim, and H. Y. Song  \textquotedblleft Autocorrelation of new generalized cyclotomic sequences of period $ p^{n} $,\textquotedblright $ \  $IEICE Trans. Fundamentals, vol.E93-A, no.11,  pp.2345-2348, 2010. 
\end{thebibliography}
\end{document}